\documentclass[preprint,onecolumn,floatfix,superscriptaddress,amsmath,amsfonts,amssymb,aps]{revtex4}

\usepackage{graphicx}
\usepackage{dcolumn}
\usepackage{bm}
\usepackage[normalem]{ulem}
\usepackage{color}
\usepackage{amssymb}
\usepackage{amstext}

\begin{document}


\title{Fingerprints of the Jahn-Teller and superexchange physics in optical spectra of manganites}

\author{N. N. Kovaleva}
\email{kovalevann@lebedev.ru}
\affiliation{P.N. Lebedev Physical Institute of the Russian Academy of Sciences, Leninsky prospekt 53, 119991 Moscow, Russia}
\author{O. E. Kusmartseva}
\affiliation{Department of Physics, Loughborough University, LE11 3TU Leicestershire, United Kingdom}
\affiliation{Mediterranean Institute of Fundamental Physics, Via Appia Nuova 31, 00040 Marino, Rome, Italy}
\author{K. I. Kugel}
\affiliation{Institute for Theoretical and Applied Electrodynamics, Russian Academy of Sciences, Moscow, Izhorskaya str. 13, 125412, Russia}
\affiliation{National Research University Higher School of Economics, Moscow, 101000, Russia}
\date{\today}

\begin{abstract}
Transition metal oxides incorporating Jahn-Teller (JT) ions Mn$^{3+}$ ($3d^4$) in manganites and Cu$^{2+}$ ($3d^9$) in cuprates exhibit outstanding magnetic and electronic properties, including  colossal magnetoresistance (CMR) and high-temperature superconductivity (HTSC). The physics of these compounds is associated with the orbitally-degenerate electronic states of JT ions, leading to the linear electron-lattice interaction, complemented by the entirely electronic superexchange (SE) interaction. Here, we discuss the fingerprints of the JT and SE physics in the temperature-dependent complex dielectric function spectra and multi-order Raman scattering spectra measured in a detwinned LaMnO$_3$ crystal, which is a famous compound exhibiting cooperative JT effect, being also a progenitor of magnetic materials exhibiting CMR effect. In accordance with the obtained experimental evidence, the associated effective parameters characterizing the JT and SE physics in manganites are presented.
\end{abstract}

\pacs{Valid PACS appear here}
\maketitle
\section{Introduction}
The rich physics of transition metal oxides incorporating Jahn-Teller (JT) ions Mn$^{3+}$ and Cu$^{2+}$ [1-17] 
possessing orbitally-degenerate $e_g$ electronic states is governed by the JT electron-lattice interaction [3,4,6,10,11], complemented by the entirely electronic superexchange (SE) interaction [1,2].  
As a rule, the former interaction results in the high-temperature JT phase transition accompanied by lifting the orbital degeneracy and the structural symmetry breaking, while the latter one paves the way for the long-range magnetic ordering and facilitates magnetic phase transitions, usually at much lower temperatures.

\begin{figure}[tbp]
\includegraphics*[width=145mm]{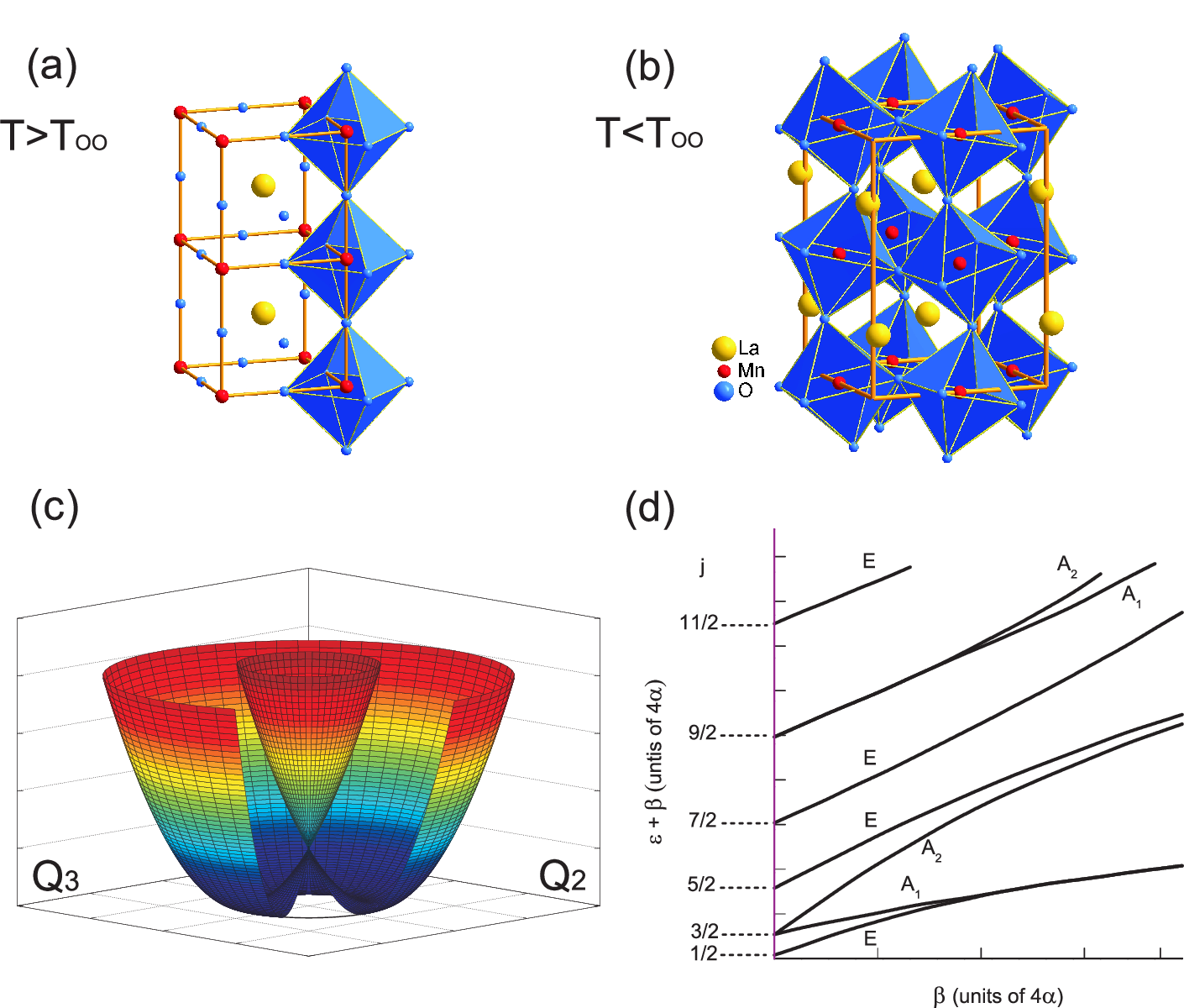}
\caption{Crystal structure of LaMnO$_3$ represented (a) by a pseudo-cubic perovskite lattice above the orbital ordering temperature $T_{OO}$\,=\,750\,K and (b) by an othorhombic $Pbnm$-symmetry lattice below the $T_{OO}$. (c) The ``Mexican hat'' APES in the normal coordinates $Q_2$ and $Q_3$ for an isolated octahedral complex [see Equation\,(\ref{MexHat})], characterized by the free rotational motion associated with the kinetic energy $E_j=\alpha j^2$, where $\vert j \vert$=1/2, 3/2, 5/2, ... are the rotational quantum numbers, $\alpha\equiv\frac{\hbar^2}{2M\rho_0^2}$, and $M$ is the mass of an octahedral apex atom. The effect of the third-order perturbation [see Equation\,(\ref{3perturb})] on the ``Mexican hat'' APES results in the formation of the three discrete minima on a circular trough at the bottom, separated by the potential barriers $\beta$ (not shown). (d) The energies of the electron-vibrational levels $\varepsilon$ in the ``Mexican hat'' APES as functions of the barrier height $\beta$, represented in the form of the diagram following the results by O`Brien \cite{O'Brien}.}
\label{fig1}
\end{figure}

Orbital degeneracy of $e_g$ electronic states in cubic environment leads to the JT instability due to the linear electron-lattice (vibronic) interaction. According to the JT theorem \cite{JT}, such a complex or a molecule, can lower its energy owing to a distortion, which lifts the degeneracy. The isolated octahedral complex can lower its energy by the three feasible tetragonal distortions along the three cubic axes. The new ground state will be lower in energy by the JT energy ($\Delta_{JT}$), possessing the infinite degeneracy, as clearly illustrated by the ``Mexican hat'' adiabatic potential energy surface (APES) (see Figure \ref{fig1}c), which is associated with the dynamic JT effect. Here, the kinetic energy due to orbital-like rotations is quantized by the angular momentum quantum numbers, $\vert j \vert$=1/2, 3/2, 5/2, ..., which can be scaled using the parameter $\alpha$ as $E_j=\alpha j^2$. However, small nonlinear perturbations can essentially affect the ground state in the ``Mexican hat'' potential relief, resulting in the formation of the three discrete minima on a circular trough at the bottom, characterized by the definite values of the $e_g$ orbital angle $\theta$ (0,\,$\pm$\,120$^\circ$ or $\pm$\,90$^{\circ}$,\,180$^{\circ}$), separated by the potential barriers of height $\beta$. The associated energies and parameters for an octahedral Cu$^{2+}$(H$_2$O)$_6$ complex based on the parameters given in the earlier study by \"{O}pik and Pryce \cite{Opik} are $\Delta_{JT}\simeq$\,0.28\,eV\,=\,3360\,K, $\alpha\simeq$\,8\,cm$^{-1}$, and $\beta\simeq$\,520\,cm$^{-1}$\,=\,65\,meV\,=\,780\,K.

In transition metal oxides represented by rare-earth $R$MnO$_3$ compounds, 
strong JT instability of the singly occupied doubly degenerate $e_g$ orbitals ($d_{x^2-y^2}$ and $d_{z^2}$) gives rise to the orbital ordering (OO) phenomena associated with collective adjustment of the distorted octahedra in the lattice structure (see Figure \ref{fig1}b), dictated by the regime of cooperative JT effect. 
In the present study we focus on LaMnO$_3$, which is a well-known famous compound exhibiting the cooperative JT effect, being also a progenitor of CMR materials. At room temperature, LaMnO$_3$ is characterized by an antiferrodistortive ordering of MnO$_6$ octahedra in orthorhombic $Pbnm$-symmetry structure, which can be associated with $C$-type OO. From the neutron diffraction study \cite{Rodriguez}, the $e_g$ orbital angle $\theta$\,$\simeq$\,107.6$^\circ$ at room temperature, being only slightly reduced to $\theta$\,$\simeq$\,106.1$^\circ$ at 573\,K. However, the OO disappears above the $T_{\rm OO}$\,=\,750\,K, which is associated with the JT transition, and for the temperatures $T_{\rm OO}\ll T \ll T^*$\,=\,1010\,K the lattice is metrically cubic \cite{Rodriguez} (as schematically presented in Figure \ref{fig1}a). Thus, the orbital angle $\theta\simeq$\,106$^\circ$--108$^\circ$ in LaMnO$_3$ crystal is somewhat reduced from the orbital angle value $\theta$\,=120$^\circ$ characteristic of an isolated octahedral complex. In addition, the magnetic phase transition occurs at the $T_{\rm N}$\,=\,140\,K, and below the $T_{\rm N}$ the Mn spins exhibit $A$-type antiferromagnetic (AFM) ordering, where the spins are ferromagnetically ordered in the $ab$ plane and antiferromagnetically stacked along the $c$ axis. The magnetic order is stabilized by the SE interactions, mediated by $d_i^4d_j^4 \rightleftharpoons d_i^5d_j^3$ virtual charge excitations between the neighboring Mn$^{3+}$ ($3d^4$) ions, involving spin and orbital degrees of freedom [2,12-14].

A successful description of the magnetic and OO phenomena in LaMnO$_3$ in a framework of the effective SE model [2,12-14] was possible [15,16] because the temperatures of orbital ordering ($T_{OO}$\,=\,750\,K \cite{Rodriguez}) and magnetic ordering ($T_{\rm N}$\,=\,140\,K \cite{Hirota}) are well separated. Here, we discuss the fingerprints of the JT and SE physics for the $e_g$ electrons in the ground state of manganites, on an example of their typical representative LaMnO$_3$, discovered in our experimental optical investigations of the Raman scattering and complex dielectric function spectra. The effective parameters associated with the JT and SE physics in LaMnO$_3$ are obtained.

\section{Samples and Methods}
LaMnO$_3$ single crystals were grown by the crucible-free floating zone method \cite{Balbashov}. The as-grown LaMnO$_3$ crystals were characterized by x-ray diffraction, energy dispersive x-ray analysis, and magnetic susceptibility measurements. 
The oxygen stoichiometry was attested by determining the N\'eel temperature in a superconducting quantum interference device (SQUID magnetometer). The AFM transition was determined at $T_{\rm N}$\,=\,139.6\,K, which is peculiar for a nearly stoichiometric LaMnO$_3$ crystal. The as-grown LaMnO$_3$ crystals have heavily twinned domain structures. We succeeded in removing the twin domains in the essential (95\%) volume fraction of the LaMnO$_3$ sample (more details are given in Ref.\,\cite{Kovaleva_PRB_LMO}). A thin slice with dimensions $\sim$3\,$\times$\,3 mm$^2$ was cut parallel to the $Pbnm$ $c$ axis and the cubic $\langle 110 \rangle$ direction. For optical measurements, the obtained surface plane was polished to optical grade . The ellipsometry measurements were carried out in the 1.2--6.0\,eV spectral range with a home-built ellipsometer of a rotating-analyzer type. For the temperature measurements (10\,--\,300\,K) the sample was mounted on a cold finger of a He-flow UHV cryostat, which was evacuated to 5\,$\times$\,10$^{-9}$\, Torr at room temperature. The ellipsometric angles $\Psi$ and $\Delta$ were measured, which are defined by the complex Fresnel reflection coefficients for light parallel ($r_p$) and perpendicular ($r_s$) to the incidence plane tan $\Psi e^{{\rm i}\Delta}=\frac{r_p}{r_s}$. The complex dielectric function spectra were extracted from the ellipsometric angles, $\Psi(\nu)$ and $\Delta(\nu)$.

The Raman scattering was studied using a Raman-microscope spectrometer LabRAM HR (Horiba Jobin Yvon) with a grating monochromator, equiped with a CCD detector. The Raman spectrometer has a spectral resolution of 0.3\,cm$^{-1}$/pixel at 633\,nm, spectral range from 100 to 4000\,cm$^{-1}$, and the unique edge filter technology. A HeNe laser (632.8\,nm) with appropriate filters was utilized under different excitation power. In the course of Raman measurements, the sample was fixed on a cold finger of the micro-cryostat cooled with liquid nitrogen. The Raman measurements were carried out in the near-normal backscattering geometry. The polarization measurements were recorded rotating the sample in the sample plane using generic linear polarization of the laser irradiation.      

\section{Experimental results and discussion}

\subsection{Theoretical spin-orbital SE model for LaMnO$_3$}
The Hamiltonian responsible for the electronic SE interactions in LaMnO$_3$, including spin and orbital degrees of freedom, is associated with virtual $d$--$d$ charge excitations, $d_i^4d_j^4 \rightleftharpoons d_i^5d_j^3$, arising from a transfer of either $e_g$ or $t_{2g}$ electrons between two neighboring Mn$^{3+}$ ions. The resulting high-spin (HS) or low-spin (LS) contributions to the SE Hamiltonian follow from the multiplet structure of the Mn$^{2+}$ ion, which depend on the intraorbital Coulomb repulsion $U$ and the Hund's exchange constant $J_H$ \cite{TanabeSugano}. In the effective SE model, the $d$--$d$ excitations between two neighboring Mn$^{3+}$ ions occur via intermediate O 2$p_{\sigma}$ orbitals and the effective hopping element implies $dd\sigma$ process \cite{ZaanenOles}. The same intersite $d$--$d$ charge excitations, which determine the SE by virtual $d_i^4d_j^4 \rightleftharpoons d_i^5d_j^3$ transitions, appear in optical spectra. 
The dominant contribution can be associated with the intersite $d_i$--$d_j$ excitations of the $e_g$ electrons, namely, 
$(t^3_{2g}e^1_g)_i(t^3_{2g}e^1_g)_j \rightleftharpoons (t^3_{2g}e^2_g)_i(t^3_{2g}e^0_g)_j$. Here, the excited states are the following \cite{Griffith}: (i) the HS ($S$=5/2) $^6A_1$ state and (ii)-(v) the LS ($S$=3/2) states $^4A_1$, $^4E$ ($^4E_{\varepsilon}$, $^4E_{\theta}$), and $^4A_2$. The excited states energies, presented via the Racah parameters $B$ and $C$ \cite{Griffith}, can be parameterized by the intraorbital Coulomb $U$ and the Hund's exchange, $J_H$=4$B$+$C$, where 4$B$\,$\simeq$\,$C$ for Mn$^{2+}$ ($d^5$) atomic values \cite{Sawatsky}. Then, the excited states energies read \cite{Oles,Khaliullin}
\begin{eqnarray}
\label{eq:E1}
{\rm(i)}\hspace{0.2cm} ^6A_1 \hspace{1cm} E_1&=&U-3J_H+\Delta_{JT},\\
\label{eq:E2}
{\rm(ii)}\hspace{0.2cm} ^4A_1 \hspace{1cm} E_2&=&U+\frac{3J_H}{4}+\Delta_{JT},\\
\label{eq:E3}
{\rm(iii)}\hspace{0.2cm} ^4E_{\varepsilon} \hspace{1cm} E_3&=&U+\frac{9J_H}{4}+\Delta_{JT}-\sqrt{\Delta^2_{JT}+J^2_H},\\
\label{eq:E4}
{\rm(iv)}\hspace{0.2cm} ^4E_{\theta} \hspace{1cm} E_4&=&U+\frac{5J_H}{4}+\Delta_{JT},\\
\label{eq:E5}
{\rm(v)}\hspace{0.2cm} ^4A_2 \hspace{1cm} E_5&=&U+\frac{9J_H}{4}+\Delta_{JT}+\sqrt{\Delta^2_{JT}+J^2_H},
\end{eqnarray}
where $\Delta_{JT}$ is the JT splitting energy between the $e_g$ levels.

In LaMnO$_3$, the $e_g$ orbitals described by the orbital angle $\theta$ alternate in the $ab$ plane, resulting in the two-sublattice OO: $\left\vert \pm\right\rangle=\cos\left(\frac{\theta}{2}\right)
\left\vert 3z^2-r^2 \right\rangle \pm \sin\left(\frac{\theta}{2}\right)
\left\vert x^2-y^2 \right\rangle$. In the associated $C$-type OO, the $e_g$ orbitals alternating in the $ab$ plane repeat along the c-axis direction. This, in turn, induces FM spin exchange between the $e_g$ orbitals in $ab$ plane and strong AFM exchange between them in the perpendicular $c$-axis direction. For each bond $\left\langle ij \right\rangle$ along a $\gamma$=$a,b,c$ direction, the Hamiltonian with coupled spin and orbital operators due to the excitations (i)--(v) reads \cite{Oles,Khaliullin}                    
\begin{multline}
H^{(\gamma)}_{ij}=\frac{t^2}{20} \left\{
-\frac{1}{E_1}(\vec S_i\cdot \vec
S_j+6)(1-4\tau_i\tau_j)^{(\gamma)}\right.
\\ \left.
+\frac{1}{8}\left(\frac{3}{E_2}+\frac{5}{E_4}\right)(\vec S_i
\cdot \vec S_j-4)
(1-4\tau_i\tau_j)^{(\gamma)}\right.   \\
+\left.\frac{5}{8}\left(\frac{1}{E_3}+\frac{1}{E_5}\right)(\vec
S_i\cdot\vec S_j-4)(1-2\tau_i)^{(\gamma)} (1-2\tau_j)^{(\gamma)}
\right\},
\label{eq1}
\end{multline}
where $t$ is an effective hopping amplitude for the $e_g$ electrons between two neighboring Mn$^{3+}$ ions and $\tau^{(\gamma)}_i$ represent the pseudospin (orbital) operators, which depend on the OO and the $\gamma$-bond direction.

For the $C$-type OO of the occupied $e_g$ orbitals, averages of the orbital projection operators can be described by the orbital angle $\theta$ as follows        
\begin{eqnarray}
(1-4\tau_i\tau_j)^{(ab)} &=& \left(\frac{3}{4} +
\sin^2
\theta\right), \\
(1-4\tau_i\tau_j)^{(c)} &=& \sin^2\theta\,, \\
(1-2\tau_i)(1-2\tau_j)^{(ab)} &=& \left(\frac{1}{2} -
\cos\theta\right)^2, \\
(1-2\tau_i)(1-2\tau_j)^{(c)} &=& (1 + \cos\theta)^2.
\end{eqnarray}

The kinetic energy $K^{(\gamma)}_n = -2 \langle H_{ij,n}^{(\gamma)}\rangle$ of an excitation with energy $E_n$ can be associated with its optical SW. The normalized optical SW can be represented in the effective number of electronic charge carriers $N_{eff}=\frac{2m}{\pi e^2N} K^{(\gamma)}_n$, where $m$ is the free electron mass and $N=a^{-3}_0=1.7\times10^{22}$\,cm$^{-3}$ is the density of Mn atoms in the cubic perovskite lattice.
For example, for the HS ($^6A_1$) excitation, the $ab$-plane and $c$-axis kinetic energies read
\begin{eqnarray}
K_{\rm HS}^{(ab)}
&=&\frac{1}{10}\frac{t^2}{E_1}{\langle\vec{S}_{i}\cdot\vec{S}_{j}+
6\rangle}^{(ab)}\;\left(\frac{3}{4}+\sin^2\theta\right),
\label{eq6}
\\
K_{\rm HS}^{(c)}
&=&\frac{1}{10}\frac{t^2}{E_1}\;{\langle\vec{S}_{i}\cdot\vec{S}_{j}+
6\rangle}^{(c)}\;\sin^{2}\theta.
\label{eq7}
\end{eqnarray}
From the above equations one can see that the temperature dependence of the kinetic energy and the associated optical SW follow the spin-spin correlations. In the classical approach, for $T\ll T_{N}$,
$\langle\vec{S}_{i}\cdot\vec{S}_{j}\rangle^{(ab)} \rightarrow$~4
and $\langle\vec{S}_{i}\cdot\vec{S}_{j}\rangle^{(c)}\rightarrow
-$4, while
$\langle\vec{S}_{i}\cdot\vec{S}_{j}\rangle^{(ab,c)}\rightarrow 0$
for $T\gg T_{N}$.
\begin{figure}[tbp]
\includegraphics*[width=150mm]{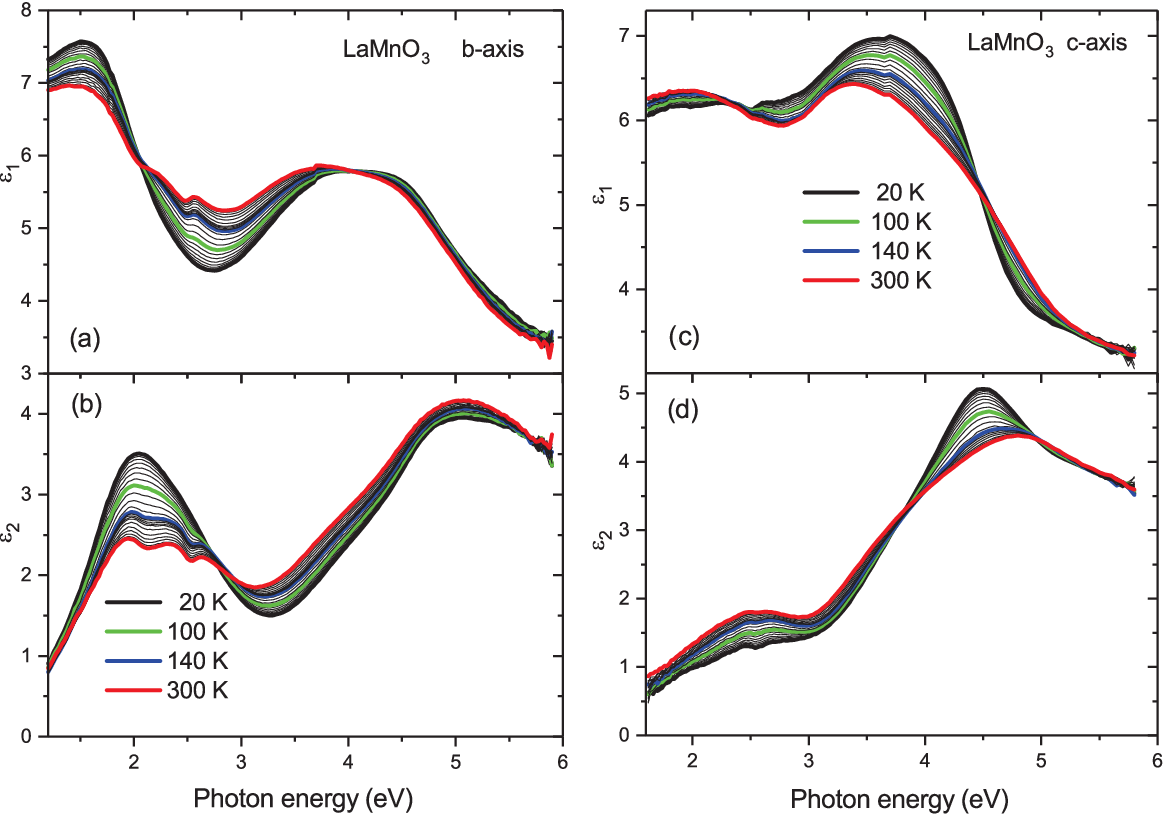}
\caption{(a) The temperature variation of real, $\varepsilon_1(\nu)$, and imaginary, $\varepsilon_2(\nu)$, parts of the complex dielectric function in LaMnO$_3$ crystal in (a) and (b) in-plane and (c) and (d) $c$-axis polarization, respectively. The spectra at the temperatures 20, 100, 140, and 300\,K around the N\'{e}el temperature $T_{\rm N}\simeq$\,140\,K are indicated by black, green, blue, and red curves, respectively. The temperature interval between two successive curves is 10\,K in the temperature range 20\,--\,200\,K and 25\,K in the temperature range 200\,--\,300\,K. Reprinted Figures 2 and 3 with permission from N. N. Kovaleva, Andrzej M. Ole\'{s}, A. M. Balbashov, A. Maljuk, D. N. Argyriou, G. Khaliullin, and B. Keimer, Phys. Rev. B {\bf 81}, 235130 (2010). Copyright 2010 by the American Physical Society.}
\label{fig2}
\end{figure}

\subsection{Fingerprits of SE physics in the experimental dielectric function spectra}
The temperature dependence of real and imaginary parts of the in-plane and $c$-axis complex dielectric function spectra, $\tilde\varepsilon(\nu)=\varepsilon_1(\nu)+{\rm i}\varepsilon_2(\nu)$, extracted from the ellipsometry measurements on a detwinned LaMnO$_3$ single crystal, are presented in Figures \ref{fig2}a and \ref{fig2}b and Figures \ref{fig2}c and \ref{fig2}d, respectively. The $\varepsilon_2$ resonance and $\varepsilon_1$ antiresonance features, appearing at the same energy, identify the available optical bands. As follows from Figure \ref{fig2}, the complex dielectric function spectra of LaMnO$_3$ are represented by two main broad optical bands located at around 2 and 5\,eV, superimposed with weaker optical features. Particularly, one can notice a fine multiplet structure appearing as a sideband to the low-energy $\sim$\,2\,eV band at higher energies. 
To obtain more accurate and detailed description on the separate contributions of different optical bands, we performed a classical dispersion analysis for the complex dielectric function in the form 
\begin{eqnarray}
\tilde\varepsilon(\nu)=\epsilon_{\infty}+\sum_j \frac{S_j\nu_j^2}{\nu_j^2-\nu^2-{\rm i}\nu\gamma_j},
\end{eqnarray}
where $\nu_j$, $\gamma_j$, and $S_j$ are the peak energy, width, and dimensionless oscillator strength of a separate Lorentz oscillator, respectively, and $\epsilon_{\infty}$ is the core contribution. First, we discuss the main trends observed in the experimental spectra. From Figure \ref{fig2}b and \ref{fig2}d, one can see that the low-energy band at $\sim$\,2\,eV is well pronounced in the room-temperature spectra in the in-plane response, whereas it is notably suppressed in the $c$-axis direction. Moreover, with decreasing temperature below the $T_{\rm N}=140$\,K, the anisotropy abruptly increases and becomes the most pronounced at low temperatures. The opposite trends, observed in the temperature dependence of the $\sim$\,2\,eV low-energy optical band, which increases in the in-plane response and decreases in the $c$-axis spectra with decreasing temperature below the $T_{\rm N}$, let us to associate it with the HS-state ($^6A_1$) transition for the $e_g$ electron transfer to an unoccupied orbital on the neighboring Mn$^{3+}$ site with a parallel spin. Indeed, the $A$-type AFM spin ordering promotes the HS-state transition in the FM $ab$ plane but suppresses it along the $c$-axis direction for the AFM aligned spins. At the same time, the higher-energy optical bands displaying the opposite trends with respect to the $\sim$\,2\,eV HS-state transition with decreasing temperature below the $T_{\rm N}$ can be associated with the LS-state ($^4A_1$, $^4E_{\varepsilon}$, $^4E_{\theta}$, and $^4A_2$) transitions, where the $e_g$ electron is transferred to an unoccupied orbital on the neighboring Mn$^{3+}$ site with an antiparallel spin. The temperature variation of the LS-state optical bands is more pronounced in the $c$-axis spectra, in particular, the optical band at $\sim$\,4.3\,eV shows noticeable intensity increase with decreasing temperature below the $T_{\rm N}$. The difference between the lowest-energy LS and HS-state transitions then defines $J_H$\,=\,0.6\,$\pm$\,0.1\,eV. The higher-energy LS-state transitions appear on a background of the strong O\,$2p$\,--\,Mn\,$3d$ charge-transfer (CT) transitions located at around $\sim$\,4.7\,eV. From a careful analysis of the temperature dependence of the complex dielectric function spectra, in combination with a classical dispersion analysis, we were able to distinguish contributions from the LS-state transitions. Then, the excited state energies in Equations (\ref{eq:E1})-(\ref{eq:E5}) read: (i) $E_1$\,=\,2.0\,$\pm$\,0.1\,eV, (ii) $E_2$\,=\,4.3\,$\pm$\,0.2\,eV, (iii) $E_3$\,=\,4.3\,$\pm$\,0.2\,eV, (iv) $E_4$\,=\,4.6\,$\pm$\,0.2\,eV, and (v) $E_5$\,=\,6.1\,$\pm$\,0.2\,eV. From this over-determined system of equations, we evaluated the consistent set of parameters: $U$\,=\,3.1\,$\pm$\,0.2\,eV, $J_H$\,=\,0.6\,$\pm$\,0.1\,eV, and $\Delta_{JT}$\,=\,0.6\,$\pm$\,0.1\,eV. 
\begin{figure}[tbp]
\includegraphics*[width=150mm]{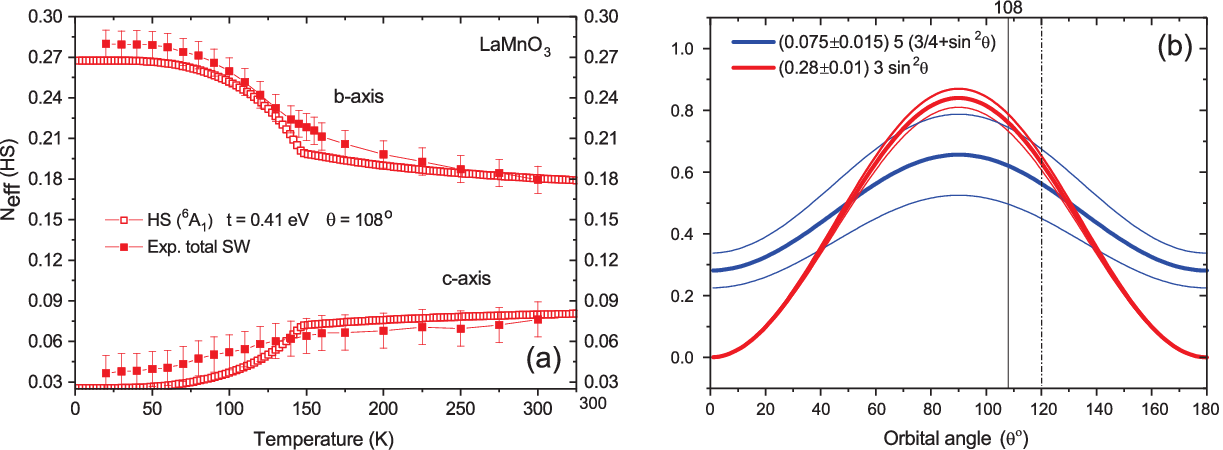}
\caption{(a) Temperature dependence of the total optical SW of the $\sim$\,2\,eV optical band, represented in terms of the effective number of charge carriers, $N_{\rm eff}$, in the in-plane and $c$-axis polarization in a detwinned LaMnO$_3$ single crystal. Reprinted Figure 11 with permission from N. N. Kovaleva, Andrzej M. Ole\'{s}, A. M. Balbashov, A. Maljuk, D. N. Argyriou, G. Khaliullin, and B. Keimer, Phys. Rev. B {\bf 81}, 235130 (2010). Copyright 2010 by the American Physical Society. (b) The $e_g$ orbital angle $\theta$ estimated from the anisotropy ratio of the low- and high-temperature optical SW of the $\sim$\,2\,eV optical band, associated with the HS-state ($^6A_1$) transition (for more details, see the text).}
\label{fig3}
\end{figure}

Figure \ref{fig3}a displays the anisotropic temperature-dependent optical SW of the $\sim$\,2\,eV low-energy optical band, $SW=\frac{\pi}{120}S_j\nu_j^2$, expressed in terms of $N_{eff}$, resulting from the classical dispersion analysis. One can see that the total optical SW of the $\sim$\,2\,eV optical band demonstrates a pronounced temperature dependence in the in-plane spectra, where the evident upturn kink below the $T_{\rm N}$\,=\,140\,K is present, whereas the relevant temperature changes are essentially suppressed in the $c$-axis spectra. As follows from the data presented in Figure \ref{fig3}a, the anisotropy ratio of the low- and high-temperature optical SW of the HS ($^6A_1$) Hubbard subband is dependent on the orbital angle $\theta$. Using Equations (\ref{eq6}) and (\ref{eq7}) in the classical approximation, one can get 
\begin{equation}
N^{(ab)}_{eff,HS} (T \ll T_{\rm N})3{\rm sin}^2\theta\,
=\,N^{(c)}_{eff,HS} (T \gg T_{\rm N})5(3/4+{\rm sin}^2\theta).
\end{equation}
 Substituting in this relation the experimentally determined limits for $N^{(ab)}_{eff,HS} (T \ll T_{\rm N})$\,=\,(0.28\,$\pm$\,0.01) and $N^{(c)}_{eff,HS} (T \gg T_{\rm N})$\,=\,(0.075\,$\pm$\,0.015) from Figure \ref{fig3}a, one can obtain an estimate for the orbital angle $\theta$, as illustrated by the graphical representation in Figure \ref{fig3}b. Here, the orbital angle $\theta$\,=\,108$^\circ$, established from the neutron diffraction experiments by Rodr\'{i}guez-Carvajal {\it et al.} \cite{Rodriguez}, is obtained for $N^{(c)}_{eff,HS} (T \gg T_{\rm N})$\,=\,0.09, within the estimated limits for $N^{(c)}_{eff,HS} (T \gg T_{\rm N})$. Using $\theta$\,=\,108$^{\circ}$, we estimate the value of the effective transfer integral $t$\,=\,0.41\,$\pm$\,0.01\,eV. It is remarkable that using the parameters estimated from our optical data, the temperature variation of the in-plane and c-axis anisotropic optical SW of the HS ($^6A_1$) Hubbard excitation calculated within the theoretical approach \cite{Oles} is well reproduced (see Figure \ref{fig3}a). In addition, the cumulative optical SW variation for the LS Hubbard excitations (ii)--(v), as well as their individual contributions, were estimated from the SE Hamiltonian [Equation (\ref{eq1})] in \cite{Kovaleva_PRB_LMO}.

\subsection{The $e \bigotimes E$ JT problem and model parameters for an octahedral complex}                         
Another important aspect to be considered here is associated with the $e \bigotimes E$ JT problem, peculiar for the singly occupied doubly degenerate $e_g$ orbitals $\left\vert x^2-y^2 \right\rangle$ and $\left\vert 3z^2-r^2 \right\rangle$ of the Mn$^{3+}$ ion. According to the JT theorem \cite{JT}, the system possessing orbital degeneracy (other than the Kramers' degeneracy) can lower its energy owing to the linear electron-lattice (vibronic) interaction, which removes the electronic degeneracy by means of the symmetry breaking structural distortion. An isolated MnO$_6$ octahedral complex with two degenerate $e_g$ orbitals may achieve lower energy by the three feasible tetragonal distortions, which can be expressed via the two independent distortions, $Q_2$ and $Q_3$, of the Mn$-$O bonds. The linear electron-lattice coupling matrix in the basis of $e_g$ electronic states $\left\vert x^2-y^2 \right\rangle$ and $\left\vert 3z^2-r^2 \right\rangle$ is 
\begin{equation}
A\left|\begin{matrix}
-Q_3 & -Q_2 \\
-Q_2 & Q_3 
\end{matrix} \right|.
\end{equation}
The APES of a JT complex includes contributions from the JT energy $\pm A \sqrt{Q_2^2+Q_3^2}$ and the elastic energy $\frac{1}{2}M\omega^2\left( Q_2^2+Q_3^2 \right)$, where $M$ is the mass of an octahedral apex atom and $\omega$ is the $Q_2$ or $Q_3$ mode circular vibrational frequency in the undistorted octahedral environment. In polar coordinates $\rho \equiv \sqrt{Q_2^2+Q_3^2}$ and $\theta \equiv {\rm tan}^{-1}\left( \frac{Q_2}{Q_3} \right)$, the adiabatic APES reads 
\begin{eqnarray}
V=\pm A \rho+\frac{1}{2}M\omega^2 \rho^2.
\label{MexHat}
\end{eqnarray}              
The APES for linear vibronic coupling in the normal coordinates $Q_2$ and $Q_3$ is double-valued as the twofold degeneracy is removed and has the so-called ``Mexican hat'' shape (see Figure \ref{fig1}c). For each distortion 
$\rho \equiv \sqrt{Q_2^2+Q_3^2}$ represented by the coupled $Q_2$ and $Q_3$ values on the circle with the radius $\left\vert \rho_0 \right\vert=\frac{A}{M\omega^2}$ one has two states at the lower and upper sheets of the APES. The new ground state is lower in energy by the JT energy $\Delta_{JT}=-\frac{A^2}{2M\omega^2}$. The infinite degeneracy of the ground state of the ``Mexican hat'' APES is achieved for all orbital angle values on the circle $\theta \equiv {\rm tan}^{-1}\left( \frac{Q_{20}}{Q_{30}} \right)$, which can be associated with the dynamic JT effect, where there are no barriers to make obstacles for radial orbital-like rotaions in $\theta$, resulting in the free of energy octahedron configuration oscillations. The kinetic energy associated with such a rotation is given by $E_j=\frac{\hbar^2}{2M\rho_0^2}j^2=\alpha j^2$, the parameter $\alpha=\frac{\hbar^2}{2M\rho_0^2}$ by definition, $\vert j \vert$=1/2, 3/2, 5/2, ... is the rotational quantum numbers, and $j=\pm\frac{1}{2}$ determines the ground state vibronic doublet.

However, the third-order (cubic) terms in the electron-lattice coupling can essentially perturb the ground state of the ``Mexican hat'' APES [5,9] 
The influence of the cubic perturbation $V_3(\theta)=A_3(\rho)\rho^3{\rm cos}3\theta$ on the lower APES was initially considered in the earlier studies [4,5]. 
It was reported that for the lower APES 
\begin{eqnarray}
V=-A\rho+\frac{1}{2}M\omega^2+A_3 \rho^3{\rm cos}3\theta
\label{3perturb}
\end{eqnarray}
this will result in the formation of the three minima at $\theta=$\,0,\,$\pm$\,120$^\circ$, if $A_3<$\,0, or $\pm$\,90$^\circ$,\,180$^\circ$, if $A_3>$\,0, which are separated by the potential barriers of height $\beta$. At a relatively low temperature compared to the barrier height, the system will be localized in one particular minimum, associated with static lattice distortions. In terms of an adiabatic model, the electron-vibrational energy levels of a Cu$^{2+}$ (3$d^9$) ion, on an example of an octahedral Cu$^{2+}$(H$_2$O)$_6$ complex embedded in a crystal, were earlier discussed  by O'Brien \cite{O'Brien}. To solve the problem of finding eigenstates and eigenvalues of the elecron-vibrational energy levels considering the third-order perturbation on the coupled electronic state $a\left\vert x^2-y^2 \right\rangle+b\left\vert 3z^2-r^2 \right\rangle$, the parameters $\alpha=\hbar/2M\rho^2$ and $\beta=-A_3\rho^3$ were introduced, and the obtained results are given in a form of the diagram \cite{O'Brien} (see also Figure \ref{fig1}d). The zero-point motion couples the excited state $\left\vert 3z^2-r^2 \right\rangle$ into the $\left\vert x^2-y^2 \right\rangle$ ground state. The order of magnitude of the $-A_3\rho^3$ effect for $\rho\simeq0.3\cdot10^{-8}$\,cm, which is regarded to be a reasonable estimate for the $\rho$ value, was obtained for Cu$^{2+}$(H$_2$O)$_6$ complex by \"{O}pik and Price \cite{Opik}. 
The wave number of zero-point oscillation is then $\omega'=\frac{3}{2\pi c}\sqrt{\frac{A_3 \rho}{M}}\simeq\frac{1350}{\sqrt{N}}$\,cm$^{-1}$, where $M=Nm_p$, $N$ is the number of protons in the ligands, and $m_p$ is the mass of the proton \cite{O'Brien}. For example, taking $N=18m_p$ for an octahedral Cu$^{2+}$(H$_2$O)$_6$ complex, this will give $\omega'\simeq$\, 320\,cm$^{-1}$. Summarizing, the values of the parameters for an octahedral Cu$^{2+}$(H$_2$O)$_6$ complex are $\rho_0\simeq0.3\cdot10^{-8}$\,cm, $\omega'\simeq$\, 320\,cm$^{-1}$, $A\simeq3.25\cdot10^{-4}$\,erg/cm, $A\rho_0\simeq4900$\,cm$^{-1}$, $\Delta_{JT}=\frac{1}{2}M\omega'^2\rho_0^2\simeq2260$\,cm$^{-1}$\,=\,0.28\,eV\,=\,3360\,K, $\alpha=\hbar/2M\rho_0^2\simeq8$\,cm$^{-1}$, $\beta=-A_3\rho_0^3\simeq520$\,cm$^{-1}$, and $A_3(\rho_0)\simeq 3.8\cdot 10^{12}$\,erg/cm$^3$.

In the case when the barriers in the trough of the ``Mexican hat'' are relatively low, the tunneling between them and fast reorientation of the octahedral configuration is nonetheless possible. 
The stationary states of a Hamiltonian for the JT system with three equivalent distortions are the following \cite{Bersuker_JETP}
\begin{eqnarray}
(\Psi_1+\Psi_2+\Psi_3)/\sqrt 3,\\
(\Psi_1-\Psi_2)/\sqrt 2,\\
(2\Psi_3-\Psi_1-\Psi_2)/\sqrt 6.
\end{eqnarray} 
Here, the tunnel splitting is 3$\Gamma$, $\Gamma<0$, so that the doubly degenerate state is the lowest. 
When the kinetic energy for the ground doublet prevail the third-order perturbation effect, the $3\Gamma$ can be approximated by $\frac{\hbar^2}{M\rho_0^2}$. 
The tunnel splitting $\Gamma \rightarrow 0$ in the limit of strong JT effect. The magnitude of $\Gamma$ can be estimated. The result depends on the higher-order perturbation terms. In a reasonable approximation, $V(\theta)\simeq \frac{1}{2}V_0(1-3{\rm cos}\theta)$, where $V_0$ is the barrier height between minima. For an effective frequency $\omega'$ for small oscillations about the equilibrium \cite{Sturge}
\begin{eqnarray}
3\Gamma\simeq\frac{3}{2\pi}\hbar \omega'{\rm exp} \left(-\frac{2V_0}{\hbar \omega'}\right)\simeq \frac{3}{2\pi}\hbar \omega' {\rm exp}\left(1-\frac{2A_3\rho^3}{\hbar \omega'}\right).
\label{3Gamma} 
\end{eqnarray}       
\begin{figure}[tbp]
\vspace{-1cm}
\hspace{3cm}\includegraphics*[width=110mm]{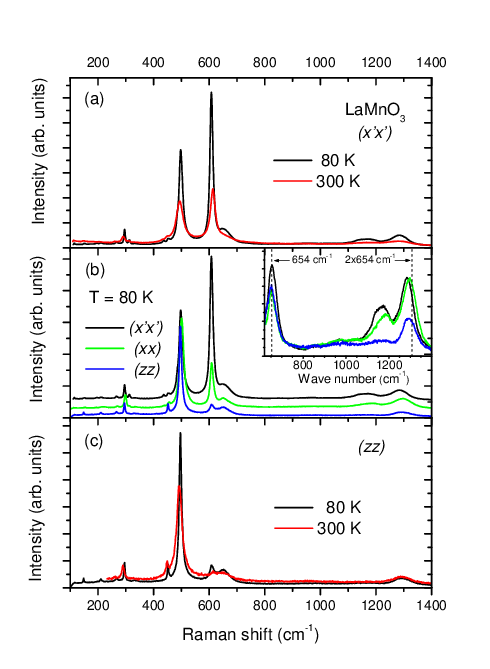}
\caption{The first- and second-order Raman scattering spectra measured on the $ac$ surface of a detwinned LaMnO$_3$ crystal at temperatures of 80 and 300\,K with the incident laser light polarized (a) at 45$^\circ$ to the principal axes and (c) along the $c$ axis. (b) The Raman spectra at 80\,K normalized to the $A_g$ mode at 495\,cm$^{-1}$. Inset: the zoom-in of the polarized first- and second-order Raman spectra, subtracted from the lattice phonons. 
Reprinted Figure 1 with permission from N. N. Kovaleva, O. E. Kusmartseva, K. I. Kugel {\it et al.}, J. Phys.: Condens. Matter. {\bf 25}, 155602 (2013). Copyright 2013 by the IOP.}
\label{fig4}
\end{figure} 
\begin{figure}[tbp]
\includegraphics*[width=160mm]{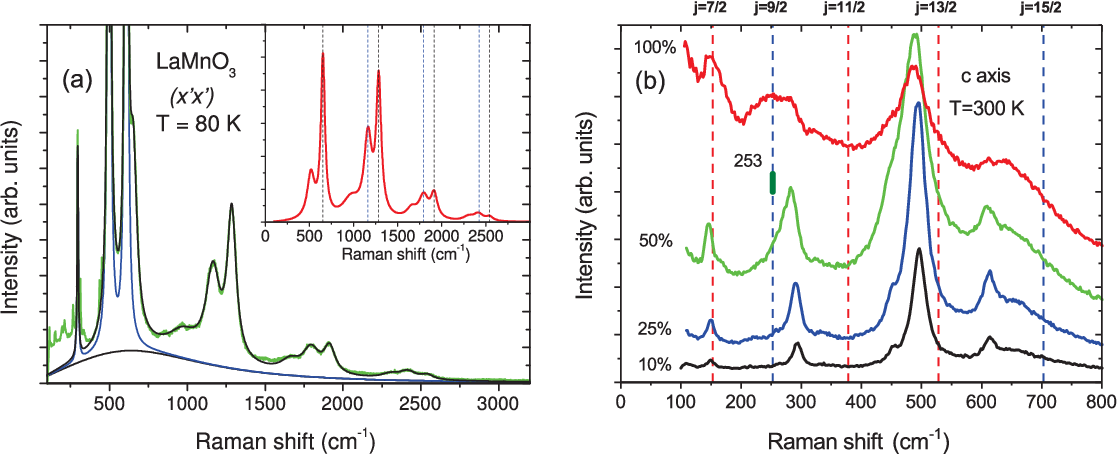}
\caption{(a) The multi-order polarized Raman spectra measured at 80\,K on the $ac$ surface of a detwinned LaMnO$_3$ crystal with the incident laser light polarized at 45$^\circ$ to the principal axes. The solid black curve represents the fitting of the experimental spectra with the Lorentzians. Inset: the fitting curve subtracted from the first-order lattice phonons and the background. The multi-order Raman scattering appears periodically up to the fourth order with a period of 630\,cm$^{-1}$, being shifted by about $\Omega$\,$\simeq$\,25\,cm$^{-1}$ from the zero frequency. Reprinted Figure 4 with permission from N. N. Kovaleva, O. E. Kusmartseva, K. I. Kugel {\it et al.}, J. Phys.: Condens. Matter. {\bf 25}, 155602 (2013). Copyright 2013 by the IOP. (b) The first-order Raman spectra measured at 300\,K in LaMnO$_3$ crystal in the $c$-axis polarization under different laser exitation power. The actual local temperatures evaluated from the FWHM of the $A_g$ phonon at 500\,cm$^{-1}$ are 10\% $\sim$ 300\,K, 25\% $\sim$ 400\,K, 50\% $\sim$ 600\,K, and 100\% $\sim$ 700\,K.}
\vspace{1.5cm}
\label{fig5}
\end{figure}

\subsection{Fingerprints of JT physics in the experimental Raman spectra}
Figures \ref{fig4}a, \ref{fig4}b, and \ref{fig4}c show first- and second-order micro-Raman spectra registered in LaMnO$_3$ crystal at 80 and 300\,K for the HeNe laser irradiation polarized along the principal crystallographic directions, as well as along the direction rotated by 45$^\circ$ in the $ac$ plane. Here, the two strongest zone-center $A_g$ (495\,cm$^{-1}$) and $B_{1g}$ (606\,cm$^{-1}$) modes correspond to the in-plane JT-type anti-symmetric and symmetric stretching of the corner-shared oxygen octahedra, respectively [17,26,27]. 
To determine temperature dependencies of the Raman modes' resonant frequencies and their full widths at half maximum (FWHM), the spectra were fitted with a set of Lorentzians. We determined linear temperature coefficients for the FWHM of the $A_g$ mode (0.07\,$\pm$\,0.01\,cm$^{-1}$/K) and $B_{1g}$ mode (0.04\,$\pm$\,0.01\,cm$^{-1}$/K), which allow to control the actual local temperatures under different laser excitation power \cite{Kovaleva_Raman_rotor}. In the first-order phonon Raman scattering, a broad band at 650\,cm$^{-1}$ is available, which cannot be associated with any normal mode, being also present in the Raman scattering spectra of other rare-earth $R$MnO$_3$ manganites at approximately the same frequency. As documented in the measured Raman polarization measurements, the 650\,cm$^{-1}$ excitation is contributed by the $A_g+B_{1g}$ symmetry modes (see Figure \ref{fig4}b). In the second order, we clearly identify three broad features representing the extended structure, where the first-order 650\,cm$^{-1}$ peak appears to be shifted by approximately 25\,cm$^{-1}$ from the zero frequency.    
Based on our investigation of the multi-order Raman scattering in LaMnO$_3$ and a comprehensive data analysis \cite{Kovaleva_Raman}, we showed that the anomalous features observed up to the fourth order are the sidebands to the low-frequency mode $\Omega=25$\,cm$^{-1}$. The triplet structure has pronounced peaks ($i=1-3$) in the $n$th-order Raman spectra ($n=1,2,3,4...$) and can be represented by the following recurrent 
formula $\omega_i^{(n)}=\Omega+\omega_i+(n-1)\omega_1$, where 
$\omega_1$\,$\simeq$\,630\,cm$^{-1}$, 
$\omega_2$\,$\simeq$\,510\,cm$^{-1}$, and $\omega_3$\,$\simeq$\,320\,cm$^{-1}$ (see Figure \ref{fig5}a and the inset). Since a sum of the phonon $k$ vectors must be $\approx$\,0, we assume that the $\omega_1$, $\omega_2$, and $\omega_3$ are the combinations from the phonons of the entire Brillouin zone which have their total $k\approx 0$. In accordance with the lattice-dynamics calculations by Iliev {\it et al.}, the $\omega_1$ feature replicated in the multi-order Raman scattering can be associated with the phonon-density-of-states (PDOS) dominated by oxygen vibrations [29,30]. 
In addition, we propose that the peak at $\omega_2$\,$\simeq$\,510\,cm$^{-1}$ is due to the contribution from the $B_{1g}$ zone-boundary phonons to the PDOS. At that the shift of the multi-order scattering by $\Omega\simeq$\,25\,cm$^{-1}$  can be associated with the electron tunneling between the minima in the ``Mexican hat'' potential. The mode $\Omega\simeq$\,25\,cm$^{-1}$ is not allowed by symmetry in Raman experiments, however, it should be allowed in the infrared experiments. Indeed, the peak at around 25\,cm$^{-1}$ is observed in the infrared spectra of rare-earh manganites GdMnO$_3$ and TbMnO$_3$ \cite{Pimenov}. 
For an isolated octahedral complex, only lightly coupled to the rest of the crystal it is 
possible to treat the JT effect introducing coupling to the rest of the crystal as a relaxation mechanism \cite{O'Brien}. However, when the coupling is strong, all the normal Raman modes must be taken into account, at that not all of them will be especially strongly coupled to the electronic state. From the relation $3\Gamma$\,=\,$\frac{\hbar^2}{M\rho^2_0}$\,$\simeq$\,25\,cm$^{-1}$ and taking $M$\,=\,16$m_p$ for an octahedral apex oxygen atom, we estimate $\rho_0\simeq0.29\cdot10^{-8}$\,cm, which is in a good agreement with the 
$\rho_0\simeq0.3\cdot10^{-8}$\,cm suggested for an octahedral complex Cu$^{2+}$(H$_2$O)$_6$ by \"{O}pik and Price \cite{Opik}. Then, using the formulae $\Delta_{JT}$\,=\,$-\frac{A\rho_0}{2}$ and $\rho_0^2=\left(\frac{A}{M\omega^2}\right)^2$, we can estimate $\Delta_{JT}$ energy, which can possibly be associated with the frequencies, $\omega_1$\,$\simeq$\,630\,cm$^{-1}$, $\omega_2$\,$\simeq$\,510\,cm$^{-1}$, and $\omega_3$\,$\simeq$\,320\,cm$^{-1}$, observed in the multi-order Raman scattering. A reasonable value of the JT energy $\Delta_{JT}$\,($\omega_3$\,=\,320\,cm$^{-1}$)\,$\simeq$\,0.25\,eV is obtained for the collective mode $\omega_3\simeq$\,320\,cm$^{-1}$. At that the zone-center JT-type $A_g$ and $B_{1g}$ modes associated with the in-phase anti-symmetric and symmetric stretching of the corner-sharing oxygen octahedra in the LaMnO$_3$ lattice are observed at higher frequencies of 495 and 606\,cm$^{-1}$, respectively [17,26,27]. 
For comparison, the associated energies $\Delta_{JT}$\,($\omega_3$\,=\,510\,cm$^{-1}$)\,$\simeq$\,0.64\,eV and $\Delta_{JT}$\,($\omega_3$\,=\,630\,cm$^{-1}$)\,$\simeq$\,0.98\,eV for the collective modes $\omega_1$ and $\omega_2$ observed in the multi-order Raman scattering are too large to be realistic. 
Using the frequency $\omega'\simeq320$\,cm$^{-1}$, we estimate the effective parameters for the JT linear electron-lattice coupling $A=\frac{2\Delta_{JT}}{\rho_0}\simeq2.76\cdot10^{-4}$\,erg/cm. Taking $3\Gamma\simeq$\,25\,cm$^{-1}$, we obtain $\beta=-A_3\rho_0^3\simeq$\,450\,cm$^{-1}$\,=\,56\,meV\,=\,670\,K and $A_3\approx-3.7\cdot10^{12}$\,erg/cm$^3$ from Equation (\ref{3Gamma}).

In addition, we investigated the polarized first- and second-order Raman scattering in LaMnO$_3$ crystal measured under different HeNe laser excitation power \cite{Kovaleva_Raman_rotor}. In the first-order spectra in the phonon spectral range we were able to identify extra features arising under the high laser excitation power, which can be associated with quantum-rotor orbital excitations in the ``Mexican hat'' APES in the dynamic limit. At the same time, the lattice phonons, corresponding to the $Pbnm$ symmetry, become almost suppressed (see Figure \ref{fig5}b). We find that additional pronounced features arise at the low phonon energies at $\sim$\,150 and $\sim$\,253\,cm$^{-1}$, which can be associated with quantum rotor orbital excitations in the dynamic limit of the JT effect, $E_j=\alpha j^2$, with the orbital momentum $j$ values of 7/2 and 9/2, respectively. The experimentally determined frequencies of the quantum rotor orbital excitations well agree with the value of the parameter $\alpha$, estimated from the tunneling frequency $3\Gamma\simeq$\,25\,cm$^{-1}$. The observed extra features appear when the local temperature was $\sim$\,700\,K, being nearly equal to the barrier height $\beta$\,$\simeq$\,450\,cm$^{-1}$\,=\,56\,meV\,=\,670\,K and slightly below the OO temprature $T_{OO}$\,$\simeq$\,750\,K, corresponding to the transition into the high-temperature pseudo-cubic phase. 

\section{Conclusions}
On one hand, our experimental study of the temperature-dependent complex dielectric function spectra in a detwinned LaMnO$_3$ single crystal, accompanied by the comprehensive data analysis, demonstrated that the $e_g$ electrons in the ground state exhibit fingerprints of the purely electronic SE interactions due to the intersite $d_i^4d_j^4 \rightleftharpoons d_i^5d_j^3$ charge excitations, including both spin and orbital degrees of freedom. The successful theoretical description in the framework of the effective SE model was possible because the temperatures of magnetic ordering ($T_{\rm N}$\,=\,140\,K) and orbital ordering ($T_{OO}$\,=\,750\,K) are well separated in LaMnO$_3$.

We have discovered that 

(i) The onset of AFM order at the $T_{\rm N}$\,=\,140\,K initiates critical optical SW redistribution between the HS and LS Hubbard subbands, disposed at around 2 and 5\,eV, respectively.

(ii) The optical SW transfer follows the magnetic ordering pattern in LaMnO$_3$: in the FM $ab$ plane the SW transfers from the high-energy LS-state excitations to the low-energy HS-state excitation, whereas in the $c$-axis polarization the SW transfers in the opposite direction.

(iii) From the description of the optical SW redistribution in the framework of the multiplet structure of the Mn$^{2+}$ ion, based on the Tanabe-Sugano diagram, the effective parameters $U$\,=\,3.1$\pm$0.2\,eV, $J_H$\,=\,0.6$\pm$0.1\,eV, $\Delta_{JT}$\,=\,0.6$\pm$0.1\,eV, and the effective hopping amplitude $t$\,=\,0.41$\pm$0.1\,eV were evaluated.

(iv) The $e_g$ orbital angle $\theta$, estimated from the low-temperature $ab$-plane optical SW and the high-temperature $c$-axis optical SW, is in a good agreement with its value $\theta$\,$\simeq$\,108$^\circ$, established from the neutron diffraction study \cite{Rodriguez}.

On the other hand, our experimental study of the polarized multi-order Raman scattering in a detwinned LaMnO$_3$ single crystal brings the evidence that the $e_g$ electrons, governed by the vibronic electron-lattice coupling effects, manifest quantum lattice effects associated with the tunneling transition between the potential energy minima arising near the JT Mn$^{3+}$ ion due to the lattice anharmonicity: 

(v) The tunneling electron transition at $3\Gamma\simeq$\,25\,cm$^{-1}$ is identified by the evidence of the low-energy mode at this energy.
 
(vi) The multi-order Raman scattering appears as sidebands, being activated by this low-energy electronic motion.

(vii) The estimated effective parameters characterizing JT physics for the $e_g$ electrons in LaMnO$_3$ are the following: $\rho_0\simeq0.29\cdot10^{-8}$\,cm, $\Delta_{JT}\simeq$\,0.25\,eV\,=\,3000\,K, $A\simeq2.76\cdot10^{-4}$\,erg/cm, $\beta\simeq$\,450\,cm$^{-1}$\,=\,56\,meV\,=\,670\,K, and $A_3\simeq-3.7\cdot10^{12}$\,erg/cm$^3$.

(viii)  The additional pronounced features popping up at low phonon energies at $\sim$\,150 and $\sim$\,253\,cm$^{-1}$ under high HeNe laser excitation power can be associated with quantum rotor orbital excitations in the dynamic limit of the JT effect, $E_j=\alpha j^2$, with the orbital momentum $j$ values of 7/2 and 9/2, respectively.

The presence of the low-energy mode at $\Omega\simeq$\,25\,cm$^{-1}$ is also confirmed by the independent experimental records [32,33]. 
The frequency of the $\Omega\simeq$\,25\,cm$^{-1}$ low-energy mode is in accord with the tunneling transition frequency 3$\Gamma \simeq$\,16\,cm$^{-1}$ in an isolated Cu$^{2+}$(H$_2$O)$_6$ complex \cite{Opik,O'Brien}.

From our optical studies of the temperature-dependent dielectric function spectra in LaMnO$_3$, we have established that the the kinetic energy variation of $e_g$ electrons from the low- to high-temperature limit, related to the optical SW of the HS-state $d_i^4d_j^4 \rightleftharpoons d_i^5d_j^3$ transition at $\sim$2\,eV, is $\sim$\,55\,meV [15,16]. 
This energy well agrees with the energy of the potential barrier height $\beta\simeq$\,450\,cm$^{-1}$\,=\,56\,meV in LaMnO$_3$, estimated in the present study. This result indicates that the SE interaction may be crucial, facilitating the tunneling transition in LaMnO$_3$ in the conditions of the cooperative JT effect. \\

{\bf Acknowledgement}\\
The authors acknowledge fruitful discussions with V Hizhnyakov. We are grateful to A Balbashov for growing the crystals. We  also thank A Kulakov for detwinning the crystal, J Strempfer, I~ Zegkinoglou, and M Schulz for characterization of the LaMnO$_3$ sample.\\


\begin{thebibliography}{99}

\bibitem{Goodenough} Goodenough J B 1995 {\it Phys. Rev.} {\bf 100} 564

\bibitem{Khomskii} Kugel K I and Khomskii D I 1982 {\it Usp. Fiz. Nauk} {\bf 136} 621 
[{\it Sov. Phys. Uspekhi} {\bf 25} 231]

\bibitem{JT} Jahn H A and Teller E 1937 {\it Proc. R. Soc. London} A {\bf 161} 220 

\bibitem{Opik}\"{O}pik U and Pryce M H L 1957 {\it Proc. Roy. Soc.} A {\bf 238} 425

\bibitem{O'Brien}O'Brien M C M 1964 {\it Proc. Roy. Soc.} A {\bf 281} 323 

\bibitem{Stoneham} Stoneham A M 1975 {\it Theory of Defects in Solids} (Oxford: Clarendon Press) 

\bibitem{BVO}Bersuker I B, Vekhter B G and Ogurtsov I I 1975 
{\it Sov. Phys. Uspekhi} {\bf 18} 569

\bibitem{KaplanVekhter} Kaplan M D and Vekhter B G  1995 {\it Cooperative Phenomena in Jahn-Teller Crystals} (New York: Plenum Press)

\bibitem{Bersuker}Bersuker I B 2006 {\it Jahn-Teller Effect} (New York, Melbourne, Cape Town, Singapore, Sao Paulo: Cambridge Univ. Press)

\bibitem{Bersuker_JETP}Bersuker I B 1963 {\it Sov. Phys. JETP} {\bf 16} 933

\bibitem{Bersuker_ChemRev}Bersuker I B
2021 {\it Chem. Rev.} {\bf 121} 1463 

\bibitem{Oles} Ole\'{s} A, Khaliullin G, Horsh P, and Feiner L F 2005 {\it Phys. Rev.} B {\bf72} 214431

\bibitem{Feiner}Feiner L F and Ole\'{s} A M 1999 Phys. Rev. B {\bf 59} 3295 

\bibitem{Khaliullin} Khaliullin G 2005 {\it Prog. Theor. Phys.} {\bf 160} 155 

\bibitem{Kovaleva_PRL_LMO} Kovaleva N N, Boris A V, Bernhard C, Kulakov A,
Pimenov A, Balbashov A M, Khaliullin G and
Keimer B 2004 {\it Phys. Rev. Lett.} {\bf 93} 147204

\bibitem{Kovaleva_PRB_LMO} Kovaleva N N, Ole\'s Andrzej M, Balbashov A M,
Maljuk A, Argyriou D N, Khaliullin G and Keimer B 2010 {\it Phys. Rev.} B {\bf 81} 235130

\bibitem{Kovaleva_Raman} Kovaleva N N, Kusmartseva O E, Kugel K I, Maksimov A A, Nuzhnyy D, Balbashov A M, Demikhov E I, Dejneka A, Trepakov V A, Kusmartsev F V and Stoneham A M 2013 {\it J. Phys.: Condens. Matter} {\bf 25} 155602   

\bibitem{Rodriguez} Rodr\'iguez-Carvajal J, Hennion M, Moussa F, Moudden A H, Pinsard L and Revcolevschi A 1998 {\it Phys. Rev.} B {\bf 57} R3189 

\bibitem{Hirota}Hirota K, Kaneko N, Nishizawa A and Endoh Y 1996 {\it J. Phys. Soc Jpn.} {\bf 65} 3736 

\bibitem{Balbashov} Balbashov A M, Karabashev S G, Mukovsky Ya M and Zverkov S A 1996 {\it J. Cryst. Growth} {\bf167} 365

\bibitem{TanabeSugano}Sugano S, Tanabe Y and Kamimura H 1970 {\it Multiplets of Transition-Metal Ions in Crystals} (New York: Academic Press)

\bibitem{ZaanenOles}Zaanen J and Ol\'{e}s A M 1993 {\it Phys. Rev.} B {\bf 48} 7197 

\bibitem{Griffith} Griffith J S 1961 {\it The Theory of Transition-Metal Ions} (Cambridge, England: Cambridge Univ. Press)  

\bibitem{Sawatsky}Zaanen J and Sawatsky G A 1990 {\it J. Solid State Chem.} {\bf 88} 8  

\bibitem{Sturge} Sturge M D 1967 {\it Sol. State Phys.} {\bf 20} 92
 
\bibitem{Iliev3}Iliev M N, Abrashev M V, Lee H-G, Popov V N, Sun Y Y, Thomsen C, Meng R L and Chu C W 1998 {\it Phys. Rev.} B {\bf 57} 2872

\bibitem{Smirnova}Smirnova I S 1999 {\it Physica} B {\bf 262} 247

\bibitem{Kovaleva_Raman_rotor}Kovaleva N N, Kusmartseva O E, Kugel K I and Kusmartsev F V,  
2017 {\it Journal of Physics: Conf. Series} {\bf 833} 012005 

\bibitem{Iliev1}Iliev M N, Abrashev M V, Popov V N and Hadjiev V G 2003 {\it Phys. Rev.} B {\bf 67} 212301 

\bibitem{Iliev2}Iliev M N, Hadjiev V G, Litvinchuk A P, Yen F, Wang Y-Q, Sun Y Y, Jandl S, Laverdi\'{e}re J, Popov V N and Gospodinov M M 2007 {\it Phys. Rev.} B {\bf 75} 064303 

\bibitem{Pimenov}Pimenov A, Mukhin A A, Ivanov V Yu, Travkin V D, Balbashov A M and Loidl A 2006 {\it Nature Phys.} {\bf 2} 97

\bibitem{Saitoh}Saitoh E, Okamoto S, Takahashi K T, Tobe K, Yamamoto K, Kimura T, Ishihara S, Maekawa S and Tokura Y 2001 {\it Nature} {\bf 41} 180

\bibitem{Gruninger}Gr\"{u}ninger M, R\"{u}kamp R, Windt M, Reutler P, Zobel C, Lorentz T, Freimuth A and Revcolevshi A 2001 {\it Nature} {\bf 418} 39

\end{thebibliography}
\end{document}